\documentclass[twocolumn,prd,superscriptaddress,floatfix,amsmath,
preprintnumbers,
showpacs,nofootinbib]{revtex4}
\usepackage{epsfig}

\usepackage[dvips]{color}
\definecolor{Black}{named}{Black}
\definecolor{Blue}{named}{Blue}
\definecolor{Red}{named}{Red}

\newcommand{\I}{{\rm i}}

\newcommand{\D}{{\rm d}}

\begin{document}

\title{Triggering collective oscillations by
three-flavor effects}

\author{Basudeb Dasgupta}
\affiliation{Max-Planck-Institut f\"ur Physik
(Werner-Heisenberg-Institut), F\"ohringer Ring 6, 80805 M\"unchen,
Germany}

\author{Georg G. Raffelt}
\affiliation{Max-Planck-Institut f\"ur Physik
(Werner-Heisenberg-Institut), F\"ohringer Ring 6, 80805 M\"unchen,
Germany}

\author{Irene Tamborra}
\affiliation{Max-Planck-Institut f\"ur Physik
(Werner-Heisenberg-Institut), F\"ohringer Ring 6, 80805 M\"unchen,
Germany}
\affiliation{Dipartimento Interateneo di Fisica ``Michelangelo Merlin,''
 Via Amendola 173, 70126 Bari, Italy}
\affiliation{Istituto Nazionale di Fisica Nucleare, Sezione di Bari,
Via Orabona 4, 70126 Bari, Italy}

\preprint{MPP-2010-12}

\begin{abstract}
Collective flavor transformations in supernovae, caused by
neutrino-neutrino interactions, are essentially a two-flavor
phenomenon driven by the atmospheric mass difference and the small
mixing angle~$\theta_{13}$. In the two-flavor approximation, the
initial evolution depends logarithmically on $\theta_{13}$ and the
system remains trapped in an unstable fixed point for
$\theta_{13}=0$. However, any effect breaking exact
$\nu_\mu$--$\nu_\tau$ equivalence triggers the conversion. Such
three-flavor perturbations include radiative corrections to weak
interactions, small differences between the $\nu_\mu$ and $\nu_\tau$
fluxes, or non-standard interactions. Therefore, extremely small
values of $\theta_{13}$ are in practice equivalent, the fate of the
system depending only on the neutrino spectra and their mass
ordering.
\end{abstract}

\maketitle

\section{Introduction}

Dense neutrino ensambles exhibit collective flavor
transformations~\cite{Pantaleone:1992eq, Sigl:1992fn,
Kostelecky:1994dt, Pastor:2001iu, Sawyer:2005jk, Duan:2005cp}. For
supernova (SN) neutrinos, these effects imprint intriguing features
on the processed spectrum such as spectral splits~\cite{Duan:2006an,
Hannestad:2006nj, Duan:2007mv, Raffelt:2007cb, Fogli:2007bk, Fogli:2008pt,
Dasgupta:2009mg, Fogli:2009rd, Duan:2007bt, Dasgupta:2008my, Duan:2007sh,
Duan:2008za, Dasgupta:2008cd}. Collective oscillations originate
from a characteristic instability in flavor space of the interacting
dense neutrino system that partly swaps its flavor content to
minimize its energy. Assuming the primary $\nu_e$ and $\bar\nu_e$
fluxes exceed those of the other flavors, the dominant effect arises
for inverted neutrino mass hierarchy. It is driven by the
atmospheric mass difference and the mixing angle $\theta_{13}$ that
is known to be small and could vanish entirely.

The transformation arising from an instability implies that the
processed spectrum is independent of the mixing angle as long as it
is small (collective transformations in the presence of matter are
suppressed for maximal mixing because one projects on the
interaction direction, but this $\cos 2\theta$ effect is irrelevant
if $\theta\ll1$ \cite{EstebanPretel:2008ni, EstebanPretel:2009ca}).
In a two-flavor treatment, $\theta$ enters only as a trigger to the
subsequent evolution, so in the SN context a very small $\theta$
shifts logarithmically the onset radius for collective
transformations \cite{Hannestad:2006nj, Duan:2007mv}. In numerical
studies, choosing $\theta$ as small as allowed by the machine
precision barely impacts the processed spectrum, although for
$\theta=0$ the system remains stuck in the unstable fixed-point
solution defined by the initial conditions.

Such a situation looks unphysical---it does not seem plausible that,
at least in principle, one can distinguish between $\theta$ being
exactly zero and some arbitrarily small but non-zero value. One may
speculate, for example, that quantum fluctuations could trigger the
transformation even for $\theta=0$ \cite{Hannestad:2006nj}, noting
that collective transformations actually preserve flavor lepton
number.

The purpose of our paper is to show that in real life we do not need
to worry about such subtleties. If $\theta_{13}$ is sufficiently
small, three-flavor effects \cite{EstebanPretel:2007yq,
Dasgupta:2007ws, Fogli:2008fj, Gava:2008rp, Blennow:2008er, 
EstebanPretel:2008ni} trigger
the instability and the logarithmic $\theta_{13}$ dependence
saturates at a small but non-zero value.

Why are collective SN neutrino transformations an effective two-flavor
phenomenon anyway? In the outer layers the temperature is too low to
support thermal $\mu$ or $\tau$ populations, obviating the possibility
to distinguish between $\nu_\mu$ and $\nu_\tau$ by charged-current
reactions. Ignoring radiative corrections, these flavors are exactly
equivalent, allowing us to define new flavors $\nu_x$ and $\nu_y$ such
that effectively $\theta_{23}=0$. If in addition $\theta_{13}=0$, one
of the new states, say $\nu_y$, becomes equivalent to $\nu_3$,
decoupling entirely from the other flavors: We are left with a
two-flavor system consisting of $\nu_e$ and $\nu_x$, governed by
$\theta_{12}$ and the solar mass difference~$\delta m^2$.

This system can show collective transformations. However, the solar
mass hierarchy is normal, suppressing the dominant transformation
effect if the primary fluxes show the usual excess of $\nu_e$ and
$\bar\nu_e$. Moreover, the collective oscillation region is at larger
radii because the solar mass difference is small, so multiple-split
effects are more easily suppressed by adiabaticity violation
\cite{Fogli:2008pt, Dasgupta:2009mg}. However, collective oscillations
driven by the solar mass difference do modify the spectra in some
scenarios~\cite{Friedland:2010sc}.

Once we allow for a small but non-vanishing $\theta_{13}$,
collective $\nu_e\leftrightarrow\nu_y$ transformations become
possible that are driven by the atmospheric mass difference $\Delta
m^2$ and occur in the usual region of large neutrino flux densities.
Our main point is that for $\theta_{13}=0$ these transformations are
triggered by small perturbations of the exact $\nu_\mu$--$\nu_\tau$
equivalence because $\nu_e$ and $\nu_x$ then no longer form an 
exact two-flavor system. Such perturbations include radiative 
corrections to the $\nu_\mu$ and $\nu_\tau$ matter 
effect~\cite{Botella:1986wy, Mirizzi:2009td}, or small 
$\nu_\mu$--$\nu_\tau$ flux differences. The
latter can be caused by the presence of muons in deeper layers of
the SN core and by radiative corrections to the interaction rates,
modifying the relative $\nu_\mu$ and $\nu_\tau$ opacities.
Non-standard interactions can also break the $\nu_\mu$--$\nu_\tau$
symmetry \cite{Blennow:2008er}, a possibility that we will not
pursue here.

We begin in Sec.~\ref{sec:EOM} with a brief discussion of the
equations of motion. In Sec.~\ref{sec:mutau} we prove that for
$\theta_{13}=0$ and for exact $\nu_\mu$--$\nu_\tau$ equivalence,
collective oscillations driven by the atmospheric mass difference
are not possible, justifying the usual two-flavor treatment. In
Sec.~\ref{sec:mutaubreak} we study concrete departures from
$\nu_\mu$--$\nu_\tau$ equivalence in the limit $\theta_{13}=0$ and
show that collective transformations are triggered by these effects.
In schematic models we compare them with an equivalent $\theta_{13}$
that would trigger collective transformations at the same onset
radius. In Sec.~\ref{sec:SN} we consider a realistic SN and study
the competition between a small $\theta_{13}$ and a small
$\nu_\mu$--$\nu_\tau$ flux difference. We conclude with a brief
summary in Sec.~\ref{sec:conclusions}.

\section{Equations of Motion}                          \label{sec:EOM}

\subsection{Matrix form}

For our conceptual discussion it is sufficient to consider the
simplest three-flavor system showing collective transformations. We
take the neutrino ensemble to be homogeneous and isotropic, study
its time evolution, and describe mixed neutrinos by matrices of
densities $\varrho_{E}$ for each energy mode $E$. We use an overbar
to represent the corresponding quantities for antineutrinos. 
Diagonal entries are the usual occupation numbers, whereas 
off-diagonal entries encode phase information. The equations of
motion (EoM) are
\begin{equation}\label{eq:eom1}
\I\,\dot{\varrho}_{E}=[{\sf H}_{E}, \varrho_{E}]
\quad\hbox{and}\quad
\I\,\dot{\bar{\varrho}}_{E}=[\bar{\sf H}_{E}, \bar{\varrho}_{E}]\;.
\end{equation}
The Hamiltonian $3\times3$ matrix is made up of the vacuum, matter,
and neutrino-neutrino terms
\begin{equation}
{\sf H}_{E}= {\sf H}^{\rm vac}_{E}+{\sf H}^{\lambda}+{\sf H}^{\nu\nu}.
\label{eq:ham}
\end{equation}
Here ${\sf H}^{\rm vac}_{E}={\sf U}{\sf M}^2{\sf U}^{\dagger}/2 E$,
with ${\sf U}={\sf R}_{23}{\sf R}_{13}{\sf R}_{12}$ the neutrino
mixing matrix and ${\sf M}={\rm diag}(m_1,m_2,m_3)$ the mass matrix.
We use the standard notation of ${\sf R}_{ij}$ as the rotation
matrix between the $i$ and $j$ mass eigenstates, with argument
$\theta_{ij}$. For antineutrinos, the vacuum Hamiltonian picks up a
relative minus sign (\hbox{$\bar{\sf H}^{\rm vac}_E=-{\sf H}^{\rm
vac}_E$}), whereas all other pieces remain identical. For oscillation
studies, we may neglect terms proportional to the identity and may
write in the mass basis
\begin{equation}
{\sf H}^{\rm vac}_{E}=\frac{1}{2E}\,
{\rm diag}\left(-\frac{\delta m^2}{2}\,,\,+\frac{\delta m^2}{2}\,,
\,\Delta m^2\right)\;.
\end{equation}
The solar mass-squared difference $\delta m^2>0$, whereas the
atmospheric one $\Delta m^2<0$ for inverted mass hierarchy (IH) and
$\Delta m^2>0$ for normal hierarchy~(NH).

The matter term, due to neutrino interactions with the charged
leptons, is in the flavor basis
\begin{eqnarray}
{\sf H}^\lambda &=& \sqrt{2}G_{\rm F}\;
{\rm diag}(N_{e},N_{\mu},N_{\tau})+{\cal O}(G_{\rm F}^2)\;,
\end{eqnarray}
where  $N_{e}$ is the net electron density (electrons minus
positrons) and similarly for the other leptons. The second-order
term is due to radiative corrections and can be non-negligible at
high densities~\cite{Botella:1986wy}. The contribution associated to
$\nu$-$\nu$ interactions is
\begin{eqnarray}
{\sf H}^{\nu\nu} =
\sqrt{2}G_{\rm F}
\int\D E\;
\left(\varrho_E-\bar{\varrho}_{E}\right) + {\cal O}(G_{\rm F}^2)\;.
\end{eqnarray}
Multi-angle effects are ignored in our isotropic system. Radiative corrections 
 can be important in dense
neutrino gases at the second order~\cite{Mirizzi:2009td}.

\subsection{New interaction basis {\boldmath$e$}--{\boldmath$x$}--{\boldmath$y$}}

Since we are concerned with a system where the $\nu_\mu$ and
$\nu_\tau$ flavors are exactly or approximately equivalent, it is
more useful to introduce new flavors $x$ and $y$ that simplify the
mixing matrix~\cite{Dasgupta:2007ws}
\begin{equation}
\begin{pmatrix} \nu_e\\ \nu_x \\ \nu_y\end{pmatrix}
={\sf R}_{23}^{\dagger}
\begin{pmatrix} \nu_e\\ \nu_\mu \\ \nu_\tau\end{pmatrix}\;.
\end{equation}
Here ${\sf R}_{23}^{\dagger}$ ``unmixes'' $\nu_\mu$ and $\nu_\tau$
with the angle $\theta_{23}$.  For $\theta_{13}=0$, $\nu_y$ is the
mass eigenstate $\nu_3$. Henceforth the interaction basis is
understood to be the $e$--$x$--$y$--basis.

This basis is useful because it explicitly removes $\theta_{23}$ 
from the formalism, if the Hamiltonian and the initial conditions 
do not distinguish $\nu_{\mu}$ and $\nu_\tau$. Naturally, the 
evolution of $\nu_e$ and $\bar\nu_e$ is independent of $\theta_{23}$
in this approximation.

\subsection{Expansion in Gell-Mann matrices}

The commutator structure of the equations of motion ensures that the
trace of $\varrho_E$ is conserved, so we may re-define them to be
traceless by subtracting a term proportional to the identity matrix
$\rm I$. The traceless part can be expanded in Gell-Mann matrices
${\sf \Lambda}_i$ with the expansion coefficients forming an
8-vector $\mathbf{X}$. Thus one can project any matrix $\sf X$ as
\begin{equation}\label{eq:bloch}
{\sf X} = {\rm tr}(\sf X)
\left(\frac{{\rm I}}{3} + \mathbf{X} \cdot \frac{\mbox{\boldmath{$\sf \Lambda$}}}{2}\right)\;,
\end{equation}
where $\mbox{\boldmath$\sf \Lambda$}$ is an 8-vector of $\sf \Lambda$
matrices. We normalize as $|{\bf X}|=2/\sqrt{3}$, corresponding to
the conventions ${\sf \Lambda}_i={\sf \Lambda}_i^\dagger$, tr$({\sf
\Lambda}_i)=0$, tr$({\sf \Lambda}_i\,{\sf \Lambda}_j)=2\delta_{ij}$,
and $[{\sf \Lambda}_i,\, {\sf \Lambda}_j]=2if_{ijk}{\sf \Lambda}_k$.
Here $f_{ijk}$ are the SU(3) structure constants, where
\begin{eqnarray}
f_{123}&=&1\nonumber\\
f_{147}&=&f_{246}=f_{257}=f_{345}=-f_{156} =-f_{367} =1/2
\nonumber\\
f_{458}&=&f_{678}=\sqrt{3}/2
\end{eqnarray}
are the non-vanishing values.

The neutrino matrices of density can now be decomposed, as in
Eq.~(\ref{eq:bloch}), in terms of an 8-dimensional polarization
vector
\begin{equation}
{\varrho}_{E} = n_{E}\,
\left(\frac{{\rm I}}{3} + \mathbf{P}_{E} \cdot \frac{\mbox{\boldmath{$\sf \Lambda$}}}{2}\right)\;.
\end{equation}
$n_E$ is the total neutrino density per unit energy interval.
Analogous expressions pertain to antineutrinos.

The different parts of the Hamiltonian can also be similarly
decomposed. The vacuum Hamiltonian is
\begin{equation}
{\sf H}^{\rm vac}_{E}=\omega_{E}
\left(\frac{\rm I}{3} + \mathbf{B} \cdot \frac{\mbox{\boldmath{$\sf \Lambda$}}}{2}\right)\;,
\end{equation}
where $\omega_{E}=\Delta m^2/(2E)$. The ``magnetic field'' is
\begin{equation}
\label{B}
\mathbf{B} = \left( \begin{array}{c}
0\\
0\\
s_{13}^2\\
S_{13}\\
0\\
0\\
0\\
-\frac{1}{2\sqrt{3}}(1+3C_{13})
\end{array}
\right) + \epsilon_\omega \left(
\begin{array}{c}
c_{13} S_{12}\\
0\\
-\frac{1}{4}C_{12} ( 3 + C_{13})\\
\frac{1}{2}  C_{12}  S_{13}\\
0\\
- s_{13} S_{12}\\
0\\
\frac{\sqrt{3}}{2} C_{12} s^2_{13}
\end{array}
\right)\ ,
\end{equation}
with $\epsilon_\omega = \delta m^2/\Delta m^2$. Moreover, we use
$s_{ij}=\sin\theta_{ij}$, $c_{ij}=\cos\theta_{ij}$,
$S_{ij}=\sin2\theta_{ij}$, and $C_{ij}=\cos2\theta_{ij}$.

Ignoring a term proportional to identity, the matter term can be
written as
\begin{eqnarray}
{\sf H}^\lambda &=& \lambda\;{\rm diag}(1,0,\epsilon_{\lambda})
=
\lambda \left(\frac{\rm I}{3} +
\mathbf{L} \cdot \frac{\mbox{\boldmath{$\sf \Lambda$}}}{2}\right)\,.
\end{eqnarray}
Here $\lambda=\sqrt{2}G_{\rm F}N_{e}$ is the effective MSW
potential. For later reference we have included $\epsilon_{\lambda}
\ll 1$, encoding radiative corrections or small
$\nu_\mu$--$\nu_\tau$ flux differences. The leptonic ``magnetic
field'' is
\begin{equation}
\label{L}
{\bf L} = \left(
\begin{array}{c}
0\\
0\\
1\\
0\\
0\\
0\\
0\\
\frac{1}{\sqrt{3}}
\end{array}
\right)
+\epsilon_{\lambda}
\left(
\begin{array}{c}
0\\
0\\
- s^2_{23}\\
0\\
0\\
- S_{23}\\
0\\
- \frac{1}{2 \sqrt{3}}(1 + 3 C_{23})
\end{array}
\right)
\ .
\end{equation}
The $\nu$-$\nu$ interaction term finally is
\begin{equation}
{\sf H}^{\nu\nu}= \mu\,
\left(\frac{\rm I}{3} + \mathbf{D} \cdot \frac{\mbox{\boldmath{$\sf \Lambda$}}}{2}\right)\;,
\end{equation}
where the effective neutrino-neutrino interaction energy is
$\mu=\sqrt{2}G_{\rm F}(N+\overline{N})$. Here
$N=N_{\nu_e}+N_{\nu_\mu}+N_{\nu_\tau}$ is the overall neutrino
density, and $\overline{N}$ for antineutrinos. The collective vector $\bf
D$ is explicitly
\begin{equation}
\mathbf{D}=\int\D E\ 
\frac{n_E\,\mathbf{P}_E -
\overline{n}_E\,\overline{\mathbf{P}}_E}{N + \overline{N}}
\;.
\end{equation}
The EoM are then
\begin{eqnarray}
\label{evol3nu} \dot{\mathbf{P}}_E &=& \left(+ \omega_E \mathbf{B} +
 \lambda\mathbf{L}  + \mu \mathbf{D}\right) \times
\mathbf{P}_E\;,\\
\label{evol3antinu} \dot{\overline{\mathbf{P}}}_E &=&
\left(- \omega_E \mathbf{B} +
 \lambda\mathbf{L}  + \mu \mathbf{D}\right) \times
\overline{\mathbf{P}}_E\;.
\end{eqnarray}
The 8-dimensional vector product is defined as $(\mathbf{a} \times
\mathbf{b})_i = f_{ijk} a_j b_k$. In this form,  the problem
resembles a set of polarization vectors ${\bf P}_{E}$ precessing
under the influence of the combined magnetic fields $\bf B$, $\bf
L$, and the mean field $\bf D$ due to all polarization vectors.

\section{Exact {\boldmath$\nu_\mu$}--{\boldmath$\nu_\tau$} equivalence}
\label{sec:mutau}

In the approximation that nothing distinguishes between the
$\nu_\mu$ and $\nu_\tau$ flavor, 2--3 mixing is physically
irrelevant and we expect that oscillations reduce to a two-flavor
problem. In fact, for $\theta_{13}=0$, no collective effects
driven by the atmospheric mass difference occur.

To prove this point we study a simplified system consisting of two
Bloch vectors, representing equal numbers of neutrinos and
antineutrinos, with the single vacuum oscillation frequency
$\omega$. For $\theta_{13}=0$ the magnetic field simplifies to
\begin{equation}
{\bf B}=\frac{2}{\sqrt{3}}\,{\bf e}_8 +
\epsilon_{\omega} \left(S_{12}{\bf e}_1 - C_{12}{\bf e}_3\right)\;,
\end{equation}
where ${\bf e}_i$ are unit vectors in the 8-dimensional
flavor space. Assuming exact $\nu_\mu$--$\nu_\tau$ equivalence
implies that $\epsilon_\lambda=0$, and therefore
\begin{equation}
{\bf L}={\bf e}_3+\frac{1}{\sqrt{3}}\,{\bf e}_8\,.
\end{equation}
Likewise, if the initial $\nu_\mu$ and $\nu_\tau$ densities are
equal, the initial polarization vectors ${\bf P}=\overline{\bf P}$ are
proportional to the same linear combination of ${\bf e}_3$ and ${\bf
e}_8$.

The static vectors ${\bf B}$ and ${\bf L}$ have components in the 1,
3, and 8 directions, whereas the only dynamical component of ${\bf
H}$, the self-term ${\bf D}$, develops an ${\bf e}_2$ component. The
EoM of the ${\bf D}$ vector derives from the difference of
Eqs.~(\ref{evol3nu}) and~(\ref{evol3antinu})
\begin{eqnarray}
\dot{\bf D}=-\epsilon_{\omega}\omega
\left[ S_{12}(P_3+\overline{P}_3) + C_{12}(P_1+\overline{P}_1)\right]\,{\bf e}_2\;.
\end{eqnarray}
In other words, the vector ${\bf H}=\omega {\bf B}+\lambda {\bf
L}+\mu\,{\bf D}$ has only components in the 1, 2, 3 and 8 direction
and thus can not mix $\nu_e$ and $\nu_y$.

The same conclusion is reached if we consider the EoM in matrix
form. The part consisting of the $e$ and $x$ flavor and the $y$
flavor form separate block matrices both for the Hamiltonian matrix
and the matrices of densities. In the $e$--$x$--$y$ basis and with
$\theta_{13}=0$, the third mass eigenstate $\nu_3$ is not admixed to
the $\nu_e$ and $\nu_x$ flavors.

\section{Broken {\boldmath$\nu_\mu$}--{\boldmath$\nu_\tau$} equivalence}
\label{sec:mutaubreak}

Even in the absence of thermal $\mu$ or $\tau$ populations the exact
$\nu_\mu$--$\nu_\tau$ equivalence is broken by several sub-leading
effects that distinguish between these flavors. In this case, the
$\nu_3$ flavor does not fully decouple from the $\nu_e$--$\nu_x$
system and collective transitions driven by the atmospheric mass
difference are inevitably triggered.

The first is provided by radiative corrections to the neutrino
matter effect where charged leptons appear in the loop. Even 
in the absence of ordinary matter, similar radiative corrections 
arise for neutrino-neutrino interactions, although the detailed 
structure of the EoM becomes more complicated~\cite{Mirizzi:2009td}. 
Since collective effects require a large density of neutrinos, 
radiative corrections and thus the breaking of $\nu_\mu$--$\nu_\tau$
equivalence are unavoidable. Finally we note that differences in 
the initial $\nu_\mu$ and $\nu_\tau$ fluxes also provides the 
required instability.

\subsection{Radiative corrections to
{\boldmath$\nu_\tau$} matter effect}

The presence of matter (i.e.~$\lambda \neq 0$) has a similar
effect as decreasing the effective mixing angle, although in detail
the dynamics is more complicated. In a frame rotating around ${\bf
L}$ there is a fast-rotating transverse $B$-field that disturbs the
system and triggers the evolution~\cite{Hannestad:2006nj}.
However, if matter effects distinguish $\nu_\mu$ and $\nu_\tau$, 
they can play a more important role in trigerring collective 
oscillations, particularly for a small mixing angle. The
largest correction is for $\nu_\tau$ where a background of ordinary
matter with baryon density $N_B$ has the same refractive effect on
$\nu_\tau$ and $\bar\nu_\tau$ that would be provided by a density of
real $\tau$ leptons ($N_\tau^{\rm
eff}=2.6\times10^{-5}\,N_B$) \cite{Botella:1986wy}. This subleading 
correction is parametrized as $\epsilon_{\lambda}$ to the usual matter 
effect $\lambda$ in Eq. (\ref{L}). 

Off-diagonal terms in the Hamiltonian generated by
$\epsilon_{\lambda}$ will mix $\nu_e$ and $\nu_y$. In the limit
$\theta_{13}\rightarrow 0$, when $\nu_3$ would otherwise have
decoupled, these terms play a role similar to $\theta_{13}$, and
recouple $\nu_3$ to $\nu_e$. We can estimate the effective
$\theta_{13}$ generated by these sub-leading matter effects by
diagonalizing the Hamiltonian instantaneously in matter as
\begin{equation}
{\sf H}_E^{\rm vac}+{\sf H}^{\lambda}=
\frac{{\widetilde{\sf U}}{\widetilde{\sf M}}^2{\widetilde{\sf U}}^{\dagger}}{2E}\;,
\label{eq:findt13}
\end{equation}
where the $\widetilde{\sf M}$ and $\widetilde{\sf U}$ denote the mass
and mixing matrix in matter. Using the standard parametrization, 
Eq.~(\ref{eq:findt13}) can be solved for the parameters of
$\widetilde{\sf U}$ and $\widetilde{\sf M}$.  We find that the matter
induced $1$--$3$ mixing is
\begin{equation}
\tan 2\widetilde{\theta}_{13}=
\frac{2\epsilon_{\lambda}\epsilon_{\omega}\lambda
  S_{12}S_{23}}{2\omega+2\epsilon_{\omega}\omega
  C_{12}+(\epsilon_{\lambda}-2)\lambda+\epsilon_{\lambda}\lambda
  C_{23}}\;,
\label{t13effl}
\end{equation}
where we ignore terms beyond the leading order in
$\epsilon_{\lambda}$.  This equation should be interpreted as
providing the critical value of $\theta_{13}$ such that if
${\theta}_{13}\lesssim\widetilde{\theta}_{13}$, the role of
$\nu_\mu$--$\nu_\tau$ equivalence breaking is more important than $\theta_{13}$
itself, and the two-flavor approximation is not valid anymore.

To demonstrate this, we consider a toy model with one Bloch vector for
neutrinos $\mathbf{P}$ and one for antineutrinos
$\overline{\mathbf{P}}$ with equal length. In the two-flavor case this
would be the simple flavor pendulum without intrinsic angular
momentum. A nonvanishing mixing angle triggers an exponential growth
of the misalignment between the force direction and the initial
orientation. The time (or distance) after which an ${\cal O}(1)$
deviation from the initial orientation is achieved grows
logarithmically with decreasing mixing angle.  We define the radius at
which there is a change of 1\% in the $\nu_e$ flavor content as the
onset radius.  In Fig.~1, we show the onset radius for this system,
and how it depends on $\theta_{13}$. We use $\lambda = 100$~km$^{-1}$,
$\epsilon_\lambda = 5 \times 10^{-5}$, $\mu=10$ km$^{-1}$,
$\omega = 1$ km$^{-1}$, and the mixing angles $\theta_{12} = 0.6$,
$\theta_{23} = \pi/4$.  Using Eq.~(\ref{t13effl}) for the chosen
parameters, we expect that the matter induced mixing becomes important
at $\theta_{13}\approx 10^{-7}$. This is in good agreement with what
we find. The logarithmic increase of the onset radius stops below this
critical mixing angle.

\begin{figure}[!t]
\label{fig1}
\includegraphics[width=3.2in]{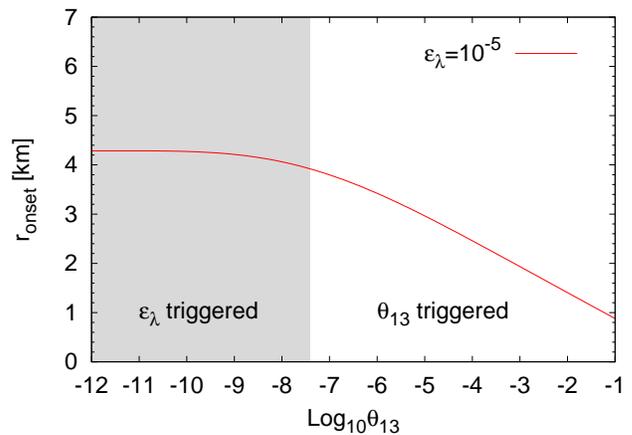}
\caption{Onset radius as a function of $\theta_{13}$ for the toy model
  described in the text. We assume a
  small difference between $\nu_\mu$ and $\nu_\tau$ refraction
  parameterized by $\epsilon_{\lambda}=10^{-5}$.}
\end{figure}

\subsection{Different primary {\boldmath$\nu_\mu$} and {\boldmath$\nu_\tau$} fluxes}

Another way to break the exact $\nu_\mu$--$\nu_\tau$ equivalence is
through an initial flux difference. Although this effect is inevitable
it has not been studied in detail. Deep in a SN core, the temperature
is large enough to support a thermal muon population, slightly
modifying the primary fluxes. Moreover, the same radiative effects
that create a refractive difference between $\nu_\mu$ and $\nu_\tau$
also modify the scattering rates and the two flavors will have
slightly different opacities and therefore different thermally driven
fluxes. Obviously the discrete nature of particle emission and
thermal fluctuations of the regions emitting the neutrinos would
necessarily make the two spectra different. 

\begin{figure}[!t]
\label{fig2}
\includegraphics[height=3.2in, angle=270]{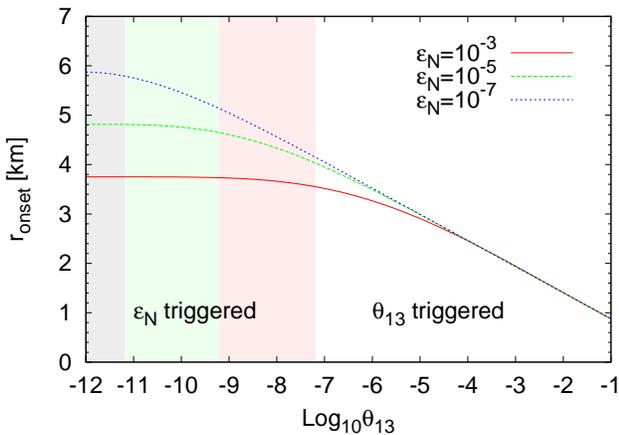}
\caption{Same as Fig.~1, now for equal $\nu_\mu$ and $\nu_\tau$
  refraction, but for different fluxes as indicated by the values of
  $\epsilon_N$.}
\end{figure}

As a toy example we again assume two equal Bloch vectors ${\bf P}$ and
$\overline{\bf P}$.  The difference between the initial densities of
$\nu_\mu$ and $\nu_\tau$ is parameterized as
\begin{equation}
\epsilon_N=\frac{N_{\nu_\mu}-N_{\nu_\tau}}{N_{\nu_e}}\;,
\end{equation}
and the same for antineutrinos. Ignoring the matter effect, we have
explicitly
\begin{equation}
\dot{\bf D}=\omega{\bf B}\times\left({\bf P}+\overline{\bf P}\right)\;,
\end{equation}
which dynamically generates components of ${\bf D}$ along ${\bf
e}_{5}$ and ${\bf e}_{7}$ even though initially ${\bf H}$ has only
components in the 1, 2, 3, and 8 directions. Therefore ${\bf H}$
develops a component along ${\bf e}_{5}$, leading to a mixing of
$\nu_e$ and $\nu_y$.

It is not straightforward to define an effective mixing angle in this
case. The effect of the different fluxes for $\nu_\mu$ and $\nu_\tau$
is to provide terms proportional to ${\epsilon_{N}\mu}$ to the
$\nu_x$--$\nu_y$ block in the Hamiltonian. These terms are themselves
dynamical (time-dependent), and are communicated to the
$\nu_e$--$\nu_y$ block by the mixing between $\nu_e$ and $\nu_x$.  The
effective mixing angle can be thought as being the initial
misalignment of ${\bf P}$ from the Hamiltonian which is approximately
proportional to $\epsilon_{N}/(\omega+\lambda)$. As this is a
three-flavor effect, it must vanish when
$\epsilon_{\omega}\rightarrow0$. We therefore expect
\begin{equation}
\widetilde{\theta}_{13}\sim
\frac{\epsilon_{N}\epsilon_{\omega}\omega}{\omega+\lambda}\;.
\label{t13effn}
\end{equation}
The logarithmic increase of $r_{\rm onset}$ with decreasing
$\theta_{13}$ saturates at $\theta_{13}$ approximately equal to the
effective mixing $\widetilde{\theta}_{13}$ induced by unequal
$\nu_\mu$--$\nu_\tau$ fluxes.

In Fig.~2, we plot $r_{\rm onset}$ for this system as a function of
$\theta_{13}$ for different values of $\epsilon_{N}$, illustrating
this effect. We use the frequencies ($\omega, \mu, \lambda) =
(1,10,100)$ km$^{-1}$, and the mixing angles $\sin^2\theta_{12} =
0.314$, and $\sin^2\theta_{23} = 0.5$. Using Eq. (\ref{t13effn}) for
the chosen parameters, we expect the flux-asymmetry induced mixing to
become important at $\theta_{13}\sim
\epsilon_{N}/(3\times10^{-3})$. This is in good agreement with what we
find. The logarithmic increase of the onset radius stops below the
estimated value of the mixing angle.

\section{Realistic Supernova}
\label{sec:SN}

We finally consider a more realistic SN example in a single-angle
treatment. The neutrinos are assumed to be emitted isotropically 
from the neutrinosphere at $R_\nu=10~{\rm km}$. 
We assume equal luminosities for all neutrino flavors, given by
\begin{equation}
L= 1.2 \times 10^{52}\  \mathrm{erg/s}\ ,
\end{equation}
and thermal spectra with average energies $\langle E_{\nu_e}\rangle=10$,
$\langle E_{\bar\nu_e}\rangle=15$, and $\langle
E_{\bar\nu_{\mu,\tau}}\rangle=20$~MeV. The electron density of the matter 
is the same as in \cite{Schirato:2002tg} at $t = 1$ s after the bounce.
For the neutrino mixing parameters we use
\begin{eqnarray}
\Delta m^2 &=& 2\times 10^{-3}\mathrm{\ eV}^2\ ,
\nonumber\\
\delta m^2 &=& 8\times 10^{-5}\mathrm{\ eV}^2\ ,
\nonumber\\
\label{theta12}
\sin^2\theta_{12}&=&0.31\ ,
\nonumber\\
\sin^2\theta_{23}&=&0.50\ .
\end{eqnarray}

With these assumptions, we have calculated the onset radius for
collective transformations as a function of $\theta_{13}$, assuming a
flux difference $\epsilon_N=10^{-5}$ and ignoring radiative
corrections to the matter effect.  Our results are shown in Fig.~3. For 
$\theta_{13} \lesssim 10^{-3}$,
 the onset radius is not sensitive to $\theta_{13}$, as expected from
Eq.~(\ref{t13effn}).

\begin{figure}[!h]
\label{fig3}
\includegraphics[height=3.2in, angle=270]{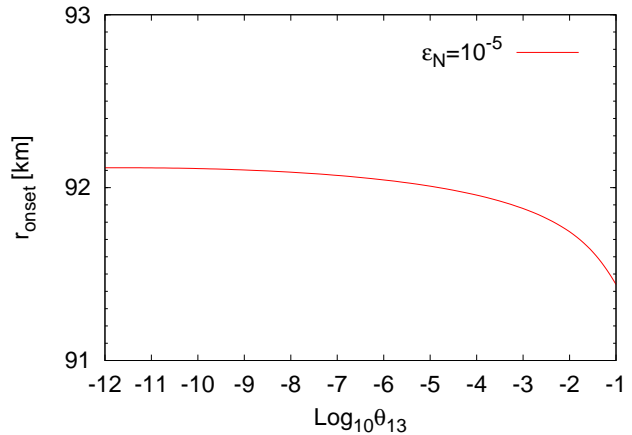}
\caption{Onset radius of collective oscillations for a realistic
SN example, assuming $\epsilon_{N}=10^{-5}$.}
\end{figure}

In a realistic SN, even when the $\nu_\mu$--$\nu_\tau$ equivalence is
perfect, the onset radius depends only very weakly on
$\theta_{13}$. The unavoidable breaking of this symmetry by radiative
corrections and the the presence of charged muons in the deep SN core
almost completely removes the $\theta_{13}$ dependence 
in the three-flavor context. Of course, MSW transitions caused by the
ordinary matter effect depend on $\theta_{13}$ in the usual way.

\section{Conclusions}                          \label{sec:conclusions}

Collective oscillations are an instability-driven phenomenon.  The
system transits from its initial unstable configuration to a stable
one, triggered by the influence of a disturbance. Usually one thinks
of this disturbance as being provided by the small offset between the
relevant flavor and the propagation eigenstates, encoded into the
mixing angle $\theta_{13}$. When this mixing angle is exactly
vanishing, one would naively think that the oscillations do not take
place.

However, one should recognize that a system sitting on an unstable 
fixed point is bound to be disturbed, unless there are symmetries that 
forbid all perturbations capable of providing an initial 
disturbance. In the neutrino oscillation context, this symmetry 
happens to be the $\mu$--$\tau$ symmetry - which is explicitly 
broken. Consequently, collective oscillations are inevitable.
This means that collective oscillations take place as usual even at 
$\theta_{13}=0$, once triggered by subleading effects.

Another fundamental point is that SN neutrino oscillations are not
sensitive to arbitrarily small values of the mixing angle.  The
fantastic sensitivity to an arbitrarily small mixing angle, as it
appears in two-flavor analyses, disappears when one takes into account
other sub-leading corrections. As a result, strategies outlined in
Refs.~\cite{Duan:2007bt, Dasgupta:2008my} may be useful for
determination of the mass hierarchy if the relevant signals are
observed, but not for determination of a non-zero $\theta_{13}$
itself. On the other hand, in principle we could determine the
neutrino mass hierarchy even if $\theta_{13}$ were exactly
zero---which might end up being our only hope if $\theta_{13}$ is
beyond the reach of laboratory-based oscillation experiments.

\section*{Acknowledgements} 

This work was partly supported by the Deutsche
Forschungsgemeinschaft under grant TR-27 ``Neutrinos and Beyond''
and the Cluster of Excellence ``Origin and Structure of the
Universe'' (Munich and Garching). The work of I.T.\ has been partly supported by the Italian
MIUR and INFN through the ``Astroparticle Physics'' research
project.  Her stay
in Munich has been partly supported by a junior fellowship
awarded by the Italian Society of Physics (Borsa SIF
``Antonio Stanghellini'').


\end{document}